\documentclass[graybox]{svmult}
\usepackage{amssymb,amsmath,natbib}
\usepackage{wasysym}
\usepackage{graphicx}
\usepackage{graphics}
\usepackage{hyperref}
\usepackage{fancybox}
\usepackage{epsfig,psfrag}
\usepackage{newlfont}
\usepackage{setspace}
\usepackage{appendix}
\usepackage{natbib}
\usepackage{color}
\usepackage{multirow}

\usepackage{mathptmx}       
\usepackage{helvet}         
\usepackage{courier}        
\usepackage{type1cm}        
\usepackage{makeidx}         
\usepackage{graphicx}        
\usepackage{multicol}        
\usepackage[bottom]{footmisc}

\makeindex             

\begin{document}

\title*{Multilevel latent Gaussian process model for mixed discrete and continuous multivariate response data}
\titlerunning{Multilevel latent Gaussian process model for mixed response data}
\author{Erin M. Schliep and Jennifer A. Hoeting}
\institute{Erin M. Schliep \at Colorado State University, Fort Collins, CO, USA,  \email{schliep@stat.colostate.edu}
\and Jennifer A. Hoeting \at Colorado State University, Fort Collins, CO, USA,  \email{jah@rams.colostate.edu}}

\maketitle

{\begin{center} \small {\bf Keywords:} mixed response data, latent variable, wetland condition, Bayesian \end{center}}

\abstract*{We propose a Bayesian model for mixed ordinal and continuous multivariate data to evaluate a latent spatial Gaussian process.
Our proposed model can be used in many contexts where mixed continuous and discrete multivariate responses are observed in an effort to quantify an unobservable continuous measurement. 
In our example, the latent, or unobservable measurement is wetland condition. 
While predicted values of the latent wetland condition variable produced by the model at each location do not hold any intrinsic value, the relative magnitudes of the wetland condition values are of interest. 
In addition, by including point-referenced covariates in the model, we are able to make predictions at new locations for both the latent random variable and the multivariate response. Lastly, the model produces ranks of the multivariate responses in relation to the unobserved latent random field. 
This is an important result as it allows us to determine which response variables are most closely correlated with the latent variable. 
Our approach offers an alternative to traditional indices based on best professional judgment that are frequently used in ecology.
We apply our model to assess wetland condition  in the North Platte and Rio Grande River Basins in Colorado. 
The model facilitates a comparison of wetland condition at multiple locations and ranks the importance of in-field measurements. }

\abstract{We propose a Bayesian model for mixed ordinal and continuous multivariate data to evaluate a latent spatial Gaussian process of wetland condition. 
Our proposed model can be used in many contexts where mixed continuous and discrete multivariate responses are observed in an effort to quantify an unobservable continuous measurement. 
In our example, the latent, or unobservable measurement is wetland condition. 
While predicted values of the latent wetland condition variable produced by the model at each location do not hold any intrinsic value, the relative magnitudes of the wetland condition values are of interest. 
In addition, by including point-referenced covariates in the model, we are able to make predictions at new locations for both the latent random variable and the multivariate response. Lastly, the model produces ranks of the multivariate responses in relation to the unobserved latent random field. 
This is an important result as it allows us to determine which response variables are most closely correlated with the latent variable. 
Our approach offers an alternative to traditional indices based on best professional judgment that are frequently used in ecology.
We apply our model to assess wetland condition  in the North Platte and Rio Grande River Basins in Colorado. 
The model facilitates a comparison of wetland condition at multiple locations and ranks the importance of in-field measurements. }

\section{Introduction}

Latent variable modeling has become common practice in a variety of scientific research fields where the latent variables are not directly observed but instead inferred from other values that are observed.
These models are particularly relevant when the observed data are assumed to be driven by some underlying, unobservable process.
Often times in the biological and ecological sciences, for example, multiple measurements are reported for each sampling unit or at each sampled location within a spatial domain and the goal is to understand the underlying latent variable(s) generating the measurements. 
Here, these measurements make up a multivariate response.
In spatial statistics, a latent variable could be used to model a random field, or process.
\cite{Chakraborty10} applied a latent spatial process model to model species abundance across a large region of South Africa.
\cite{Christensen02} developed a general framework for multivariate latent variable models that incorporates spatial correlation among the latent variables.

We focus on ordered categorical, or ordinal data where measurements for each observation are reported on a specified scale, (e.g., low, medium, high).
Some discrete data are ordinal in nature.
For example, in survey data, respondents are asked to characterize their opinions on a Likert scale ranging from strongly disagree to strongly agree.
In other situations, data will be ordinal when a researcher reports the response as a discretized continuous variable instead of as the actual continuous variable due to constraints on the data collection process.
This may be the case when reporting sediment size in streams or surface area of leaves on individual plants, especially when the data are to be collected over a large spatial domain.

We propose a model for drawing inference about mixed ordinal and continuous multivariate response data.
We refer to the model as a multilevel latent process model because we introduce latent variables at two levels within the hierarchy. 
The first level of latency is introduced by assuming there is a continuous latent process that generates each variable of the multivariate response.
The model extends the multivariate latent health factor model proposed by \cite{Chiu10} by allowing dependence on the site effect to vary across response variables.

The second level of latency is introduced by assuming there exists an underlying univariate latent spatial process, or latent random field, that is generating the multivariate response.
We assume a linear relationship between each of the latent continuous response processes (first level of latency) and the latent spatial process (second level of latency). 
Refer to Figure \ref{Diagram} for a diagram of the multilevel latency.
This model provides estimates of the latent spatial process in order to compare different locations within a specified region of interest.
Second, the model allows quantification of the relationship between the spatial latent variable and each of the variables of the multivariate response.
Lastly, we can determine which of the variables of the multivariate response are most closely associated with the latent spatial process.
In doing so, we can establish weights for each of the response variables to be used in weighted averaging for estimating the underlying latent spatial process.
By incorporating point-referenced covariate information, we can predict the value for the latent spatial variable as well as the mixed ordinal and continuous multivariate response at new locations.

In Section \ref{motivate} we motivate the model with an application of assessing the condition of wetlands in Colorado.
In Section \ref{sec:Model} we introduce the mixed ordinal and continuous multivariate latent Gaussian process model; we also describe methods of inference and estimation of the model parameters under the Bayesian framework.
In Section \ref{sec:Eval} we develop methods to predict the latent random variable and ranking procedures for the multivariate response.
The methodology is applied in Section \ref{sec:Wetlands} through the evaluation of wetland condition in the North Platte and Rio Grande River Basins of Colorado.
Section \ref{sec:Disc} concludes with a brief discussion and recommendations for future work.

\section{Motivating example}
\label{motivate}

The proposed model was motivated by a program to asses the condition of wetlands in Colorado. 
Limited data exist on the location, type, and condition of Colorado's wetlands hindering wetland management. 
The long-term viability and integrity of Colorado's wetland resources are threatened due to increased demand from major urban areas for water development and storage projects, growth in the oil and gas industry, and changes in forest health \citep{Dahl2011}.
The data considered here were collected through a partnership between Colorado Parks and Wildlife (CPW)'s Wetlands Program and the Colorado Natural Heritage Program (CNHP) to assess the condition of wetlands in Colorado.  
The specific data used in this model were collected in Colorado's North Platte and Rio Grande River Basins \citep{Lemly2011, Lemly2012}. 
One of the major goals of the CPW-CNHP partnership is to model the spatial distribution of wetland ecological condition throughout each river basin in the state. 
Our goal was to improve spatial  modeling techniques in order to help land managers effectively maintain and improve critical wetland habitats.

In order to implement effective wetland protection strategies and to establish restoration and management plans, wetlands must be assessed and then potential threats or stressors identified.
There are many different in-field measurements, known as metrics, that reflect various aspects of wetland condition.
These metrics can be of any variable type including continuous, count data, ordinal, etc. 
Overall scores that are computed based on multiple measurements are referred to as multi-metric indices. 
When the metrics are of the same variable type, one index to evaluate overall wetland condition is an average metric score. 
However, difficulty arrises when trying to compute an index that encompasses metrics of different variable types. 
In this work, we propose using continuous latent variables as consistent measures across all metric types.
Appropriate link functions can map the continuous latent variables to the different metrics. 

One popular index that incorporates 12 metrics to evaluate ecological condition is the index of biotic integrity, or IBI \citep{Karr81}.
It is of great interest to ecologists to determine whether the particular metrics that are used in computing the IBI are useful in evaluating wetland condition.
Of equal importance, ecologists are interested in identifying which of the measurements taken during in-field data collection are most representative of overall wetland condition.
This is beneficial as it will not only increase accuracy in gauging wetland condition but will also save time and resources for future data collection by requiring fewer measurements.

There are tens of thousands of acres of reported wetlands in Colorado's North Platte and Rio Grande River Basins and sampling time and resources are limited.
One of the major goals of the wetland profiling project is to model the spatial distribution of the ecological condition of wetlands throughout the basins and determine the optimal metrics for measuring key habitat features for wetland-dependent wildlife species.
We compare the ecological condition of the wetlands based on five metrics in both the North Platte and Rio Grande River Basins.

\section{Model and inference}
\label{sec:Model}

\subsection{Multivariate mixed response data}

\begin{figure}[h!]
\begin{center}
\hspace{-1.5cm}
\includegraphics[height=6in]{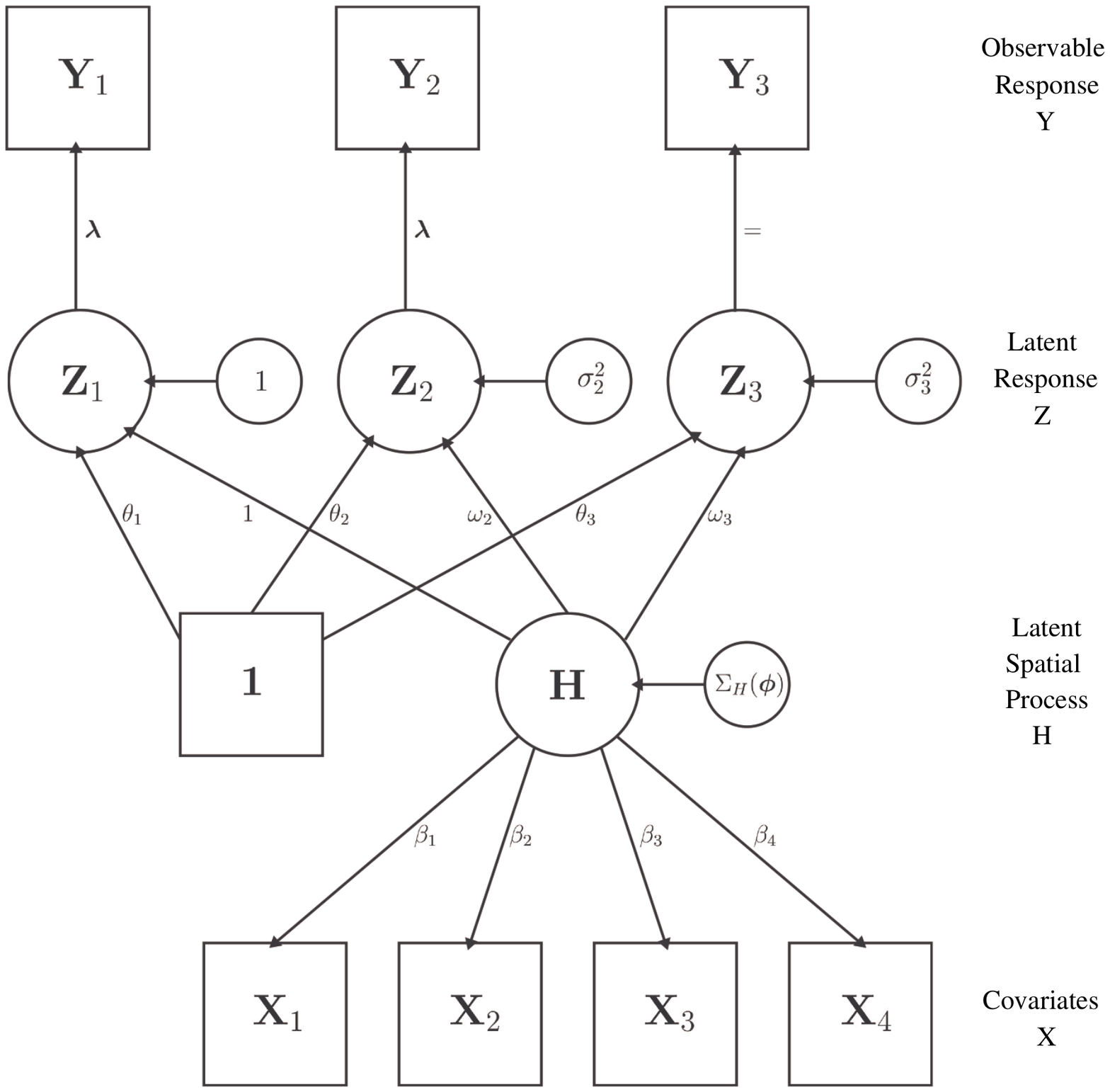}
\vspace{-4.5cm}
\caption{Diagram of a multilevel latent model with two ordinal observable response variables $\mathbf{Y}_1$ and $\mathbf{Y}_2$ and a continuous observable response variable $\mathbf{Y}_3$. Here, $\mathbf{Z}_1, \mathbf{Z}_2,$ and  $\mathbf{Z}_3$, represent the first level of latency as the latent continuous response variables. $\mathbf{H}$ is the second level of latency and is the latent spatial random field of interest. There are 4 covariates in the model, $\mathbf{X}_1, \mathbf{X}_2, \mathbf{X}_3$, and $\mathbf{X}_4$. The $\square$ indicates an observable value and the $\Circle$ indicates a random variable. Additional parameters are shown next to the links. \label{Diagram}}
\end{center}
\end{figure}

One of the main goals of this work is to use observed mixed ordinal and continuous multivariate responses from a finite number of point-referenced locations to draw inference on an underlying latent spatial process.
We wish to make predictions of the latent spatial process as well as quantify uncertainty. 
The model consists of first representing each of the multivariate response variables as a continuous response. 
For the ordinal response variables, this continuous response is latent. 
We then define a linear relationship between each of the (latent) continuous response variables and the underlying latent spatial process of interest. 
We assume that each of the response variables contains information about this latent spatial process. 
Refer to Figure \ref{Diagram} for a diagram of the multilevel latent model. 

For the spatial domain of interest, $D$, define $\left\{ \mathbf{Y}(\mathbf{s}) = [Y_1(\mathbf{s}), \dots, Y_J(\mathbf{s})], \mathbf{s} \in D \right\} $ as a mixed ordinal and continuous multivariate random field at location $\mathbf{s}$ having $J$ responses.
Each response at location $\mathbf{s}$, $\left \{Y_j(\mathbf{s}), \mathbf{s} \in D\right \}$ for $j = 1,\dots , J$ is modeled by a random field of either continuous or ordinal values.
Let  $J_c$ denote the number of continuous response variables and $J_o$ denote the number of ordinal response variables, where $J_o \ge 1$.
Therefore, $J=J_o + J_c$.
For all ordinal variables variables $j$ in $1, \dots, J_o$, the observable response $Y_{j}(\mathbf{s}) \in \{1, \dots, K\}$ for every location $\mathbf{s}$.
The model can easily be generalized to include observable response variables with varying number of categories, e.g. $Y_{j}(\mathbf{s}) \in \{1, \dots, K_j\}$.
In such a case, parameter constraints, discussed below, will need to be modified to maintain model identifiability.

We assume there exists an underlying continuous multivariate Gaussian process, $\{ \mathbf{Z}(\mathbf{s}) = [Z_{1}(\mathbf{s}), \dots, Z_J(\mathbf{s})], \mathbf{s} \in D\}$, over the region of interest that is generating $\mathbf{Y(s)}$.
Dropping the dependence on $\mathbf{s}$ for ease of notation, we denote $\mathbf{Y} = [\mathbf{Y}_1, \dots, \mathbf{Y}_J]$ and $\mathbf{Z} = [\mathbf{Z}_1, \dots, \mathbf{Z}_J]$ where $\mathbf{Y}_j$ and $\mathbf{Z}_j$ are the $j^{th}$ observable response and underlying continuous Gaussian process, respectively.
For $j = 1, \dots, J$, we define $F_j$ as the mapping of the continuous variable $\mathbf{Z}_j$ to the observable response $\mathbf{Y}_j$. 
Whereas the observable response data presented in this work are continuous and ordinal, the model holds for other types of response variables, e.g. binary, Poisson, etc.
The mapping function $F_j$ can take on any form as long as it is reasonable to assume that an underlying continuous Gaussian process is generating the response.   
For an ordinal response, the continuous variable $\mathbf{Z}_j$ is latent. 
Here, the mapping $F_j$ is defined as a function with parameter vector $\boldsymbol{\lambda}_j$, a $(K+1) \times 1$ dimension vector of thresholds, that assigns the latent continuous random variables $\mathbf{Z}_j$  into the ordered categories $1, \dots, K$ of the observable data $\mathbf{Y}_j$ \citep{Muthen1984}.
The threshold parameter vector is constrained such that  $- \infty = \lambda_{j,0} \le \lambda_{j,1} \le \dots \lambda_{j,K} = \infty$ for each ordinal metric.
We define a mapping, $F_j$, of $Z_{j}(\mathbf{s})$ to $Y_{j}(\mathbf{s})$ as
\begin{eqnarray}
Y_{j}(\mathbf{s})=F_j (Z_{j}(\mathbf{s}),\boldsymbol{\lambda}_j ) = \sum_{k=1}^K k I_{\{\lambda_{j,k-1} < Z_{j}(\mathbf{s}) \le \lambda_{j,k}\}}, \hspace{.1in} j=1, \dots ,J_o, \hspace{.05in} \mathbf{s} \in D.
\label{eqn:Y}
\end{eqnarray}
For continuous response variables, the mapping $F_j$ is taken as the identity function since $\mathbf{Z}_j$ would be observed directly.

\subsection{Multilevel latency}
We assume that the latent random process is expressed by a mixed model.
For the $j^{th}$ random process, $\mathbf{Z}_j$, we assume a multivariate Gaussian process where
\begin{eqnarray}
\mathbf{Z}_{j} \sim GP (\theta_j \mathbf{1} +  \omega_j \mathbf{H}, \sigma^2_j \mathbf{I}).
\label{eqn:Z}
\end{eqnarray}
We define the mean of each $\mathbf{Z}_{j}$ as a metric-specific linear combination of the 1-vector and a latent random field $\mathbf{H}$.
The latent random field $\mathbf{H}$ is the process of interest and encompasses the latent measure of wetland condition in our application.
The fixed effect $\theta_j$ is the intercept for metric $j$ and the fixed effect $\omega_j$ is the factor loading of the spatial random field $\mathbf{H}$.
Both $\boldsymbol{\theta}$ and $\boldsymbol{\omega}$ are $1 \times J$ dimensional vectors.
The parameter $\boldsymbol{\omega}$ allows us to quantify the relationship between each of the response variables and $\mathbf{H}$. 
The variance of $\boldsymbol{Z}_j$ is specific to each metric $j$, which we define as $\sigma^2_j \mathbf{I}$ where $\mathbf{I}$ is the identity matrix.
For $j \ne l$, $\mathbf{Z}_{j}$ and $\mathbf{Z}_{l}$ are conditionally independent given $\mathbf{H}, \boldsymbol{\theta}$, and $\boldsymbol{\omega}$.

The spatial dependence of the multivariate random field is modeled through the latent spatial process, $\mathbf{H}$.
Note that the inclusion of the additional latent process $\mathbf{H}$ makes this a multilevel latent process model.
We assume this latent spatial process is driving the mixed ordinal and continuous multivariate observable response, $\mathbf{Y}$.
Therefore, $\mathbf{H}$ provides a univariate summary measure for each location from which we will draw inference across space. 
We assume $\mathbf{H}$ to be a Gaussian process with covariates in the mean structure and a covariance matrix defined by a spatial correlation function.
Let 
\begin{equation}
\label{Hdist}
\mathbf{H} \sim GP (\boldsymbol{X \beta, \Sigma_H(\phi)})
\end{equation}
where $\mathbf{X}$ contains $p$ location-specific observable covariates and $\boldsymbol{\beta}$ is a $p \times 1$ vector of coefficients.
The covariance matrix $\boldsymbol{\Sigma_H(\phi)}$ is described by a function $\boldsymbol{\Sigma_H(\phi)} =  \rho (||\mathbf{s}_i - \mathbf{s}_l||; \boldsymbol{\phi})$ where $\rho$ is a covariance function with parameters $\boldsymbol{\phi}$ that produces a valid covariance matrix depending only on the spatial distance matrix.
\vspace{-1cm}
\vspace{-1cm}
\subsection{Bayesian framework}
The observed multivariate data matrix $\mathbf{y}$ is of dimension $n \times J$ where $n$ is the number of point-referenced locations in our sample and $J$ is thel number of metrics or responses at each location. 
For $i = 1, \dots, n$ and ordinal response variables $j = 1, \dots, J_o$, the density of $y_{ij}$ is the integral from $\lambda_{j,y_{ij}-1}$ to $\lambda_{j,y_{ij}}$ of the normal distribution defined for $Z_{ij}$. 
Whereas we first defined $\mathbf{Z}_j$ as a Gaussian process for each $j = 1, \dots, J$, realizations of these processes have a multivariate normal distribution. 
Denoting the multivariate ordinal observed response $\mathbf{y}_o = [\mathbf{y}_{1}, \dots, \mathbf{y}_{J_o}]$, we write the likelihood of the $j^{th}$ vector of $\mathbf{J}_o$, $\mathbf{y}_{j}$, as the integral of an $n$-dimensional multivariate normal distribution. 
Therefore
\begin{eqnarray} 
\begin{split}
p_o(\mathbf{y}_{j} | \mathbf{H}, \theta_j, \omega_j, \boldsymbol{\lambda}_j, \sigma^2_j) &= \int_{\lambda_{j,y_{1j}-1}}^{\lambda_{j,y_{1j}}} \cdot \cdot \cdot \int_{\lambda_{j,y_{nj}-1}}^{\lambda_{j,y_{nj}}} (2\pi)^{-n/2} |\sigma^2_j \mathbf{I}_n|^{-1/2}\\
& \hspace{-1.5cm} \times \exp \left\{- \frac{1}{2} [\mathbf{Z}_{j} - (\theta_j \mathbf{1} + \omega_j \mathbf{H})]^{'} [\sigma^2_j \mathbf{I}_n]^{-1}  [\mathbf{Z}_{j} - (\theta_j \mathbf{1} + \omega_j \mathbf{H})]\right\} d\mathbf{Z}_{j}.
\label{eqn:Ylike}
\end{split}
\end{eqnarray}

\noindent For the multivariate continuous observed response $\mathbf{y}_c = [\mathbf{y}_{1}, \dots, \mathbf{y}_{J_c}]$ , the likelihood of the $j^{th}$ vector of $\mathbf{y}_c$ is the multivariate normal density, $p_c$.
Denoting  $\mathbf{y}= [\mathbf{y}_o,\mathbf{y}_c]$, the likelihood for all observations is given by
\begin{eqnarray*} 
p (\mathbf{y} | \boldsymbol{H, \theta, \omega, \lambda, \sigma}^2) = \prod_{j=1}^{J_o} p_o (\mathbf{y}_{j} | \mathbf{H}, \theta_j, \omega_j, \boldsymbol{\lambda}_j, \sigma^2_j) \times 
\prod_{j=1}^{J_c} p_c (\mathbf{y}_{j} | \mathbf{H}, \theta_j, \omega_j, \sigma^2_j)
\end{eqnarray*}
We define prior distributions for all model parameters and latent random variables to complete the Bayesian model specification. 
We aim to assign proper yet vague prior distributions to unknown parameters to maintain generality of the model.
When applicable, conjugate priors are assigned to ease computational complexity. 

To ensure identifiability of the intercept parameter vector $\boldsymbol{\theta}$,  it is necessary to place a restriction on one of the threshold parameters. 
Where the lower and upper cut points are defined as $\lambda_{j,0} = -\infty$ and $\lambda_{j,K} = \infty$, we assume without loss of generality that $\lambda_{j,1} = 0$ for $j = 1, \dots, J_o$. 
Therefore, we are left to estimate $J_o \times (K-2)$ threshold parameters.
A uniform prior can be assigned to the cut parameters as shown in \cite{AlbertChib93}, where 
$p(\lambda_{j,k} | \lambda_{j,k-1}, \lambda_{k,k+1}) \propto I_{(\lambda_{j,k-1},\lambda_{j,k+1})}$, for $k = 2, \dots, k-1$ and $j = 1, \dots, J_o$.
However, the constraint that $\lambda_{j,k-1} \le \lambda_{j,k}$ can lead to poor mixing in the Markov Chain. 
We transform the parameter $\lambda_{j,1}, \dots, \lambda_{j,k-1}$ to a new space with parameters $\alpha_{j,1}, \dots, \alpha_{j,k-1}$ \citep{AlbertChib97}.
The transformation is performed by setting $\alpha_{j,1} = \lambda_{j,1} =0$, $\alpha_{j,2} = \log(\lambda_{j,2})$, and letting $\alpha_{j,k}=\log(\lambda_{j,k} - \lambda_{j,k-1})$ for $k = 3, \dots, K-1$.  
The inverse transformation is expressed as $\lambda_{j,k} = \sum_{i=2}^k e^{\alpha_{j,i}}$.
We then impose an unrestricted multivariate normal prior distribution to the $(K-2) \times 1$ dimension vector $\boldsymbol{\alpha}$ for each $j = 1, \dots, J_o$  with mean $\mathbf{a}$ and covariance matrix $\mathbf{A}$.

As denoted above, each of the latent response vectors $\mathbf{Z}_{j}$ for $j = 1, \dots, J$ is a Gaussian process with mean $\theta_j \mathbf{1} + \omega_j \mathbf{H}$ and covariance matrix $\sigma^2_j \mathbf{I}$.
Due to the multivariate multilevel latent structure of the model, some parameters will be fixed to ensure identifiability of the other parameters of interest. 
When the threshold vectors are metric-specific, as shown in (\ref{eqn:Y}), the scale parameter, $\sigma^2_j$, for $j = 1, \dots, J_o$ of the covariance of the continuous multivariate random variables $\mathbf{Z}_j$ will have to be fixed \citep{SkrondalRabe04}.
When all of the ordinal metrics have the same number of categories, the threshold parameter vector $\boldsymbol{\lambda}$ can be assumed the same across all metrics. 
In this case, the parameter $\boldsymbol{\sigma}^2$ is identifiable for the ordinal metrics if just one element, $\sigma^2_j$ is fixed. 
Fixing thresholds to be equal for all metrics is not overly restrictive when the number of categories of the ordered response is small.
Indeed, it can be helpful when some of the metrics have few responses in some categories. 
Also, the mean and variance of the latent continuous response are able to vary across metrics which allows the model to be flexible. 
However, this assumption becomes more restrictive as the number of categories per metric increases because the model may not be sufficiently flexible to preserve the proportions in each category for the different metrics.
Without loss of generality, we set the variance of the first ordinal response variable, $\sigma^2_1=1$ and drop the metric dependence on the thresholds. 
The remaining parameters, $\sigma^2_j$ for $j = 2, \dots, J$, are assigned inverse-Gamma prior distributions with hyper-parameters $a_z$ and $b_z$.

The mean of the distribution of the latent process $\mathbf{H}$ is defined as $\mathbf{X} \boldsymbol{\beta}$, where the covariate matrix $\mathbf{X}$ is centered and scaled and does not include the one vector in order to estimate $\boldsymbol{\theta}$ in (\ref{eqn:Z}). 
The conjugate prior distribution for the $p \times 1$ vector $\boldsymbol{\beta}$ is
$MVN(\mathbf{0}, \sigma^2_\beta \mathbf{I}_p)$.
Let $\boldsymbol{\Sigma_H(\phi)}$ be the covariance of the distribution of $\mathbf{H}$ where the vector $\boldsymbol{\phi}$ represents the parameters of the covariance function. 
Here we choose an exponential covariance function and write $\rho(\mathbf{s}_i - \mathbf{s}_l; \boldsymbol{\phi}) = \phi_1 \exp^{-d_{il} \phi_2}$ where $d_{il}$ represents the Euclidean distance between locations $i$ and $l$. 
The conjugate inverse-Gamma prior distribution is assigned to $\phi_1$ and a Gamma prior distribution is assigned to $\phi_2$. 
The shape and scale hyper-parameters of these distributions are $a_{\phi_1}$ and $b_{\phi_1}$ and $a_{\phi_2}$ and $b_{\phi_2}$, respectively. 
For identifiability, however, $\phi_1$ is set to 1 when all response variables are ordinal. 
Specification of the prior distribution of $\phi_2$ and its corresponding hyper-parameters can be challenging and must be chosen with careful consideration to keep it non-informative. (see e.g., \citealp{Schmidt2008}).

The parameters $\boldsymbol{\theta}$ and $\boldsymbol{\omega}$ are each assigned a multivariate normal prior distribution with mean vector $\mathbf{0}$ and covariance matrix $\sigma^2 \mathbf{I}_J$. 
The scale parameters of both covariance matrices, $\sigma^2_\theta$ and $\sigma^2_\omega$, are chosen to be large such that the prior distributions are vague.  
To ensure identifiability of the model parameters one value of the $1 \times J$ dimension vector $\boldsymbol{\omega}$ must be fixed. 
Without loss of generality we set $\omega_1 = 1$. 
Fixing $\omega_1$ establishes a point of reference for the relationship between $\mathbf{Z}$ and the parameter of interest, $\mathbf{H}$.

\subsection{Inference}
\label{sec:Inf}

We make inference about the parameters of the model using the Bayesian paradigm incorporating Gibbs and Metropolis Hastings sampling techniques.
This approach allows estimation of both the model parameters and the multilevel and multivariate latent variables, as well as their uncertainty.
Due to the constrained threshold parameter vector $\boldsymbol{\lambda}$, the model proposed in this work is computationally complex.

The joint posterior distribution of the unknown parameters of interest and the latent variables given the observed data can be factored and written as
\begin{eqnarray*}
\begin{split}
p(\mathbf{Z}\boldsymbol{, \theta, \omega,}\mathbf{H}\boldsymbol{, \lambda, \sigma}^2 \boldsymbol{, \beta, \phi | y}) & \propto p(\boldsymbol{y |} \mathbf{Z} \boldsymbol{, \theta, \omega,} \mathbf{H,}\boldsymbol{\lambda, \sigma}^2 \boldsymbol{, \beta, \phi}) p(\mathbf{Z} |\boldsymbol{\theta, \omega,} \mathbf{H}\boldsymbol{, \lambda, \Sigma, \beta, \phi}) \\
& \times p(\boldsymbol{H}|\boldsymbol{\beta,\phi}) p(\boldsymbol{\theta, \omega},\boldsymbol{\lambda, \sigma}^2\boldsymbol{, \beta, \phi})
\end{split}
\end{eqnarray*}
where $p(\boldsymbol{y | \cdot})$ is the distribution of the mixed ordinal and continuous multivariate random variables given the model parameters and latent variables, $p(\mathbf{Z | \cdot})$ is the conditional distribution of the continuous latent random variable, $p(\mathbf{H}|\boldsymbol{\beta, \phi})$ is the distribution of the latent spatial field of interest, and $p(\boldsymbol{\theta, \omega, \lambda, \sigma}^2 \boldsymbol{, \beta, \phi})$ is the joint prior distribution for the parameters
$\boldsymbol{\theta}, \boldsymbol{\omega}, \boldsymbol{\lambda}, \boldsymbol{\sigma}^2, \boldsymbol{\beta}$, and $\boldsymbol{\phi}$.

The Markov chain Monte Carlo (MCMC) algorithm proceeds as follows:
\begin{enumerate}
\item Update the spatial covariance scale and range parameters, $\phi_1$ and $\phi_2$, respectively. $\phi_1$ can be drawn drawn directly from its complete conditional distribution whereas $\phi_2$ requires a Metropolis-Hastings step to sample from its complete conditional distribution.
\item Update the regression parameter vector  $\boldsymbol{\beta}$ and the latent spatial multivariate normal, $\mathbf{H}$, from their complete conditional distributions.
\item Update the metric-specific parameters $\boldsymbol{\theta}$ and $\boldsymbol{\omega}$ and variance parameter $\boldsymbol{\sigma}^2$ each in block form from their complete conditional distributions.
\item Update the threshold parameters, $\boldsymbol{\lambda}$ by drawing $\boldsymbol{\alpha}$ from $p(\boldsymbol{\alpha}|\mathbf{y}_o,\mathbf{Z}_o)$ and inverse mapping to get $\boldsymbol{\lambda}$. See \cite{Higgs10} for explicit details on the reparameterization and updating scheme for $\boldsymbol{\lambda}$.
\item Update the latent multivariate normal $\mathbf{Z}_o$ from the complete conditional distribution.
\end{enumerate}
The samples from the posterior distribution can then be used to draw inference on both the model parameters and latent variables.

\section{Posterior inference}
\label{sec:Eval}
\subsection{Posterior prediction}
\label{sec:Pred}
The model can be used to make predictions for the mixed ordinal and continuous multivariate response as well as the underlying latent spatial process at unobserved locations.
The multivariate response at $m$ unobserved locations will be denoted $\widetilde{\mathbf{Y}} = [\widetilde{\mathbf{Y}}_{1}, \dots, \widetilde{\mathbf{Y}}_{J}]$ where $\widetilde{\mathbf{Y}}_{j} = [\widetilde{Y}_{1 j}, \dots, \widetilde{Y}_{m j}]'$.
Similarly, predictions of the latent spatial process at the $m$ unobserved locations will be written as $\widetilde{\mathbf{H}} = [\widetilde{H}_{1}, \dots, \widetilde{H}_{m}]'$.
Predictions can be made using the Bayesian posterior predictive distributions $p(\widetilde{\mathbf{Y}}| \mathbf{y})$ and $p(\widetilde{\mathbf{H}} | \mathbf{y})$ for the multivariate response and latent spatial process, respectively.

In most applications, the value of the latent variable $H_i$ at location $i$ will be inconsequential but the comparison of $\mathbf{H}$ across locations may be of interest.
For example, wetland condition encompasses many variables. 
If a latent variable $H_i$ summarizes wetland condition at site $i$, comparisons among sites will be useful to many agencies and individuals.
For each location, we obtain draws from the distributions $H_i | \mathbf{y}$ and $\widetilde{H}_i|\mathbf{y}$ for each iteration of the Markov chain. 
We then examine the distribution of the posterior ranks for each location to draw inference and conduct comparisons across the region of interest.

Other model parameters of particular interest include the parameters of the latent spatial field $\mathbf{H}$, $\boldsymbol{\beta}$ and $\boldsymbol{\phi}$, as well as the metric-specific parameters of $\mathbf{Z}, \boldsymbol{\omega}$, and $\boldsymbol{\sigma}^2$.
Estimating the parameter vector of coefficients of the linear model, $\boldsymbol{\beta}$, enables us to evaluate the relationship between the point-referenced covariates and the latent random variable $\mathbf{H}$.
The unaccounted for spatial correlation of the latent random variable $\mathbf{H}$ can be estimated by drawing inference on $\phi_1$ and $\phi_2$ as well as the effective range, $3/\phi_2$.
The effective range is the distance at which the correlation function does not exceed 0.05 times the variance.

\subsection{Multivariate correlation statistics}
\label{sec:Corr}
We estimate the relationship between latent variables $\mathbf{Z} = [\mathbf{Z}_1, \dots, \mathbf{Z}_J]$ and $\mathbf{H}$ by computing multiple correlation values.
Due to the deterministic relationship between latent $\mathbf{Z}$ and observed $\mathbf{Y}$, we assume that the relationship we are estimating will capture that of the relationship between $\mathbf{H}$ and the multivariate response $\mathbf{Y}$. 
This is a method used in canonical correlation analysis to evaluate the level of linear relationship between two sets of variables \citep{Rencher2002}.
It is useful to first partition the covariance matrix of the matrix $\mathbf{Z}$ and vector $\mathbf{H}$ as
\[\mathbf{S} = \left( 
\begin{array}{cc}
\mathbf{S}_{ZZ}&\mathbf{S}_{ZH}\\
\mathbf{S}_{HZ}&\mbox{s}_{HH}
\end{array}
\right)
\]
where $\mathbf{S}_{ZZ}$ is the $J \times J$ sample covariance matrix of $\mathbf{Z}$, $\mathbf{S}_{ZH}$ is the $J \times 1$ matrix of sample covariances between $\mathbf{Z}$ and $\mathbf{H}$, and $\mbox{s}_{HH}$ is the sample covariance of $\mathbf{H}$. 
The $(j, j')$ element of $\mathbf{S}_{ZZ}$ is the covariance between the $n \times 1$ dimension vectors $\mathbf{Z}_j$ and $\mathbf{Z}_{j'}$.
Similarly, the $j^{th}$ element of $\mathbf{S}_{ZH}$ is the covariance between the $n \times 1$ dimension vectors $\mathbf{Z}_j$ and $\mathbf{H}$. 
A measure of association between $\mathbf{Z}$ and $\mathbf{H}$ as a whole is $R^2_M = |\mathbf{S}_{ZZ}^{-1}\mathbf{S}_{ZH}\mathbf{S}_{HH}^{-1}\mathbf{S}_{HZ}|$. 
This value is analogous to $R^2$ in linear regression. 
This value can also be expressed in terms of the canonical correlations between $\mathbf{Z}$ and $\mathbf{H}$.
However, we would like to evaluate the correlation between each of the responses and $\mathbf{H}$ separately.
The correlation between $\mathbf{Z}_j$ and $\mathbf{H}$ is defined as the square root of
\begin{eqnarray}
R^2_{Z_j|H} = \frac{\left(\frac{1}{\mbox{s}_{HH}}\right) (\mathbf{S}_{ZH})_j}{\mbox{diag}(\mathbf{S}_{ZZ})_j}
\label{eqn:R}
\end{eqnarray}
where $(\mathbf{S}_{ZH})_j$ is the $j^{th}$ element of the $J \times 1$ vector $\mathbf{S}_{ZH}$. 

We evaluate the multiple correlation for each metric using the posterior simulations.
Therefore, at each simulation draw of the model parameters, we first compute the covariance matrix $\mathbf{S}$.
Then, for $j = 1, \dots, J$, we compute the correlation between the posterior draw of $\mathbf{Z}_j$ and $\mathbf{H}$ using (\ref{eqn:R}). 
Larger values of $R_{Z_j|H}$ (i.e., closer to 1) suggest that metric $j$ is more correlated with the underlying latent variable $\mathbf{H}$. 
In application, a large $R_{Z_j|H}$ value means that metric $j$ is a good measurement or predictor for the unobserved latent spatial process.
We use the multiple correlation values to rank the importance of each of the response metrics in measuring the latent spatial process of wetland condition.

\subsection{Model evaluation}
\label{sec:ModEval}
Mixed ordinal and continuous multivariate response models present a unique problem for model evaluation. 
Whereas there are multiple methods to measure predictive ability for discrete response models or continuous response models, the difficulty arises when we wish to compare mixed response models with both continuous and discrete variables. 
Multicategory loss functions like those presented in \cite{Higgs10} cannot be applied when $J_c \ne 0$. 
Therefore, we direct our attention to loss functions for continuous data since we have a continuous latent variable for all $J$ metrics. 
In the Bayesian framework, the loss is computed by comparing the true value to draws from the posterior predictive distribution.
Therefore, we first need to determine the ``true" value for the ordinall variable on the continuous scale.
The posterior mean or median of the latent continuous response could be used as the ``true" value but we feel this favors the discrete response metrics. 
We propose setting the ``true" value for the continuous representation of the observed ordinal variable $y$ as the value $\hat{Z}$ such that 
\begin{eqnarray}
\frac{\int_{\lambda_{y-1}}^{\hat{Z}} \frac{1}{\sqrt{2\pi \sigma^2_z}} \exp^{\frac{-1}{2\sigma^2_z}(Z-\mu_z)^2}dZ}{\int_{\lambda_{y-1}}^{\lambda_{y}} \frac{1}{\sqrt{2\pi \sigma^2_z}} \exp^{\frac{-1}{2\sigma^2_z}(Z-\mu_z)^2} dZ}=0.50
\label{eqn:fitZ}
\end{eqnarray}
where $\mu_z$ and $\sigma^2_z$ are the mean and variance of the posterior distribution of $Z$, respectively. 
Therefore, $\hat{Z}$ is the $50^{th}$ percentile of the estimated normal distribution between the thresholds $\lambda_{y-1}$ and $\lambda_y$. 
We can estimate both $\mu_z$ and $\sigma^2_z$ for $i = 1, \dots, n$ and $j = 1, \dots, J_o$ using the posterior draws of the parameters $\omega_j, \theta_j, H_i$ and $\sigma^2_j$.
We apply this method to perform model comparison in Section \ref{sec:Wetlands} under squared error loss.

\section{Assessing wetland condition}
\label{sec:Wetlands}

\subsection{Data and model specification}
\begin{figure}
\begin{center}
\includegraphics[height=2in]{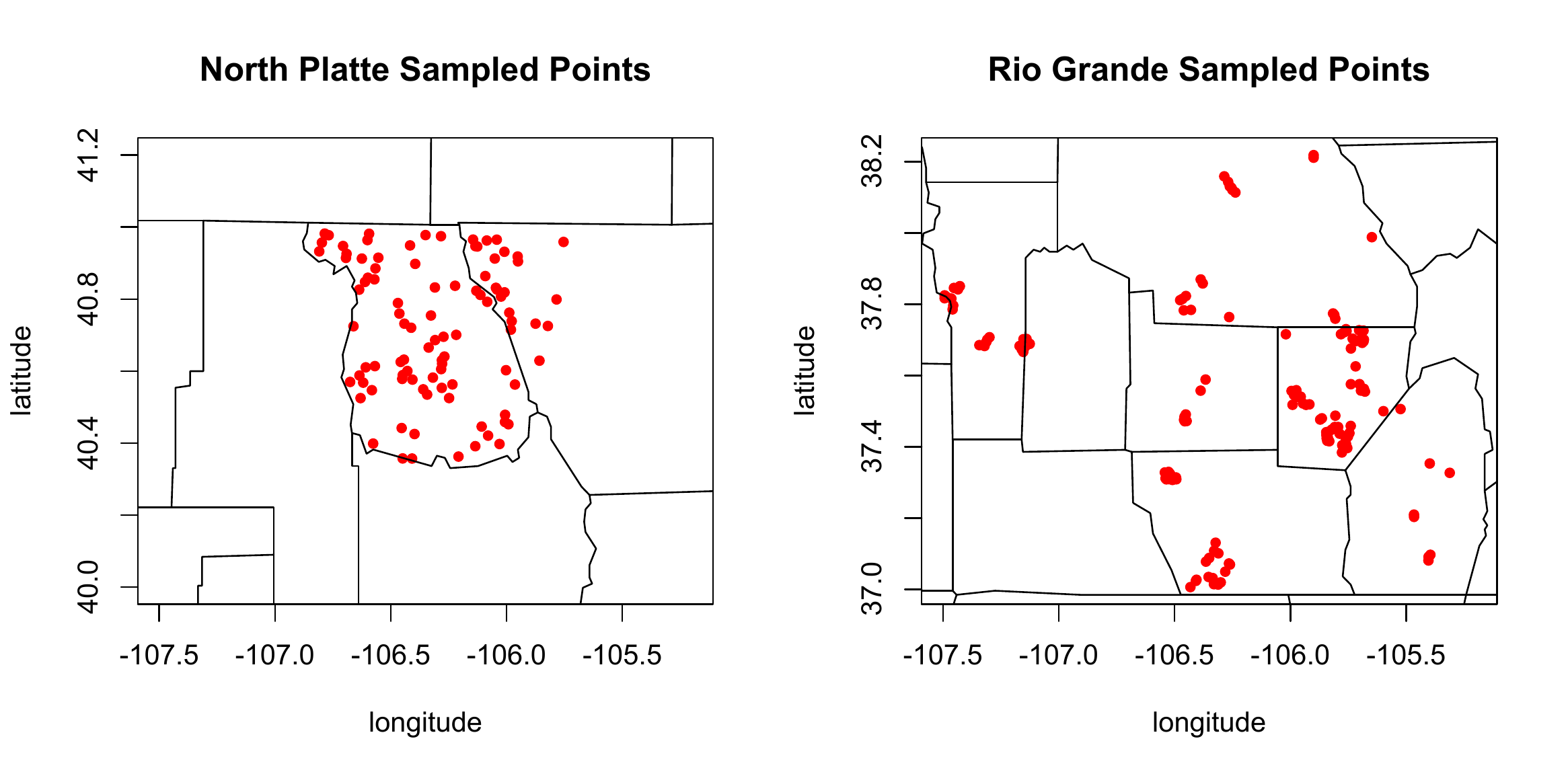}
\caption{The $n=232$ locations of observed data within the North Platte and Rio Grande River Basins.}
\label{fig:locations}
\end{center}
\end{figure}

The data were collected at 95 locations within the North Platte River Basin and 137 locations within the Rio Grande River Basin, resulting in $n=232$ locations (Figure \ref{fig:locations}).
The surveyed parcel consisted of a 0.5 hectare area around each target location.
These locations were sampled randomly using a Generalized Random Tessellation Stratified (GRTS) survey design \citep{Stevens04}.
Details of the GRTS design differed between the basins \citep{Lemly2011, Lemly2012}.
We applied the multivariate multilevel latent Gaussian process model to each river basin separately and to the basins together and reached similar conclusions.
The results presented here are those from the river basins modeled together as one data set.

The data include measurements to evaluate the biotic integrity of the wetland, as well as the surrounding landscape, soil, and water conditions. 
Here we apply our multilevel latent model to evaluate the biotic integrity of wetlands. 
We refer to the biotic integrity as a proxy for wetland condition because it's the biotic condition that drives the overall condition of the wetland. 
Five measurements, or metrics, were derived from detailed vegetation surveys conducted at each field location.
The five metrics include native plant cover, noxious weed cover, aggressive native cover, structural complexity, and floristic quality assessment \citep{Lemly2012}.
It is assumed that each of these metrics represents a component of the biotic integrity of the wetland.
It is current practice for wetland condition assessment to use a method of weighted averages to evaluate the biotic condition using these metrics.
Whereas these weights are often thought to be assigned based on best professional judgement or without statistical support, our goal is to use the data within the multivariate multilevel latent Gaussian process  model to rank the metrics in a hierarchy of most important to least important to assess wetland condition.
We can then identify a subset of the metrics that are most valuable for future data collection.

Each metric was reported on a five-category ordinal scale from ``poor" to ``excellent," to which we assign integer values from 1 to 5, respectively (Appendix \ref{data}).
The floristic quality assessment, native plant cover, noxious weed cover, and aggressive native cover metrics are discretized continuous variables (See \cite{Lemly2012} for more details on discretization). 
The floristic quality metric evaluates the overall floristic quality and fidelity of the plant community at each location to natural, or undisturbed, conditions \citep{Rocchio2007}.
Each species in the Colorado flora has been assigned a coefficient of conservatism (C value: 0-10) that reflects the species tolerance of intolerance to disturbance \citep{Swink1994, Taft1997}.
The continuous value is an average of C values assigned to the plants present at the wetland site. 
The ordinal value at each location is assigned by applying a threshold to the continuous metric value.
However, this thresholding scheme is dependent on wetland type because the natural vegetation differs between wetland type with some naturally containing plant species with lower values of floristic quality.
Structural complexity is Likert-like and has no tangible underlying continuous variable. 
Here, we fit a discrete-only model with $J_o = 5$ and $J_c=0$ as well as a mixed response model with $J_o=4$ and $J_c = 1$ where the continuous metric is floristic quality and compare the results. 
For all $J_o$ ordinal responses, the observed value $Y_i  \in \{1, \dots, K=5\}$ for $i = 1, \dots, 232$.

The variance of $Z_{i1}$ is fixed and held constant across all locations at $\sigma^2_1=1$ for model identifiability.
The hyperparameters of the inverse-gamma distributions of the metric specific variance parameters $\sigma^2_j$ are $a_z=b_z=1$ for $j = 2, \dots, 5$.
The metric specific parameters $\boldsymbol{\theta}$ and $\boldsymbol{\omega}$ are of dimension $1 \times 5$.
We set the variance hyperparameters $\sigma^2_\theta$ = $\sigma^2_\omega$ = 100.
For identifiability of the coefficient vector $\boldsymbol{\beta}$, we fix $\omega_1 = 1$.

Elevation and percent of closed tree canopy vegetation are two continuous point-referenced covariates used to model the mean of the Gaussian process $\mathbf{H}$ (Appendix \ref{data}). 
We also included wetland type as a categorical covariate with five levels: riparian shrublands and woodlands, saline wetlands, marshes, wet meadows, and fens.
The prior distribution of the coefficient vector $\boldsymbol{\beta}$ is $MVN(\boldsymbol{0}, \sigma^2_\beta \mathbf{I_p})$ with $\sigma^2_\beta = 100$ and $p=6$.
The exponential covariance function for the latent random variable $\mathbf{H}$ is defined as $\phi_1 \exp^{-d_{il} \phi_2}$ where $d_{il}$ is the Euclidean distance between locations $i$ and $l$.
In the mixed response model, we assign an $\mbox{Inv.Gamma} (1,1)$ prior for $\phi_1$ and fix $\phi_1=1$ in the discrete-only model.
In both models, $\phi_2$ is assigned a $\mbox{Gamma} (2, 2)$ prior distribution.
The prior of $\phi_2$ was chosen such that the effective range, $3/\phi_2$, could reach the maximum distance between sites.

\subsection{Model results}
\label{ModelRes}
The Markov chain Monte Carlo algorithm was run for 100,000 iterations using R software \citep{R}.
The first 10,000 iterations for both models were discarded as burn-in.
We ran multiple chains from different starting values to evaluate convergence of our Gibbs sampler. 
The Gelman \citeyearpar{Gelman2004BDA} potential scale reduction factor for each parameter was below 1.2. 
Similarly, other standard diagnostics showed no indications of lack of convergence. 

The posterior estimates from both the discrete-only model and the mixed response model indicate that wetland condition scores are higher for locations at higher elevations and with higher percentages of closed tree canopy (Table \ref{table:estimatesD} for discrete-only response model, Table \ref{table:estimatesDC} for mixed response model).
The coefficients $\beta_3, \beta_4, \beta_5,$ and $\beta_6$ represent the effect for saline, marsh, wet meadow, and fen wetland types, respectively, relative to riparian shrublands and woodlands.
These values vary greatly between models due to the discretization of the floristic quality assessment metric.
The discretization process includes additional information about the condition of each site based on its wetland type and thus, the ordinal values for this metric are not uniformly assigned across all locations \citep{Lemly2012}.
For example, a riparian wetland with a floristic quality value of 5.6 on the continuous scale would be assigned a 4 on the ordinal scale, whereas a marsh wetland with the same continuous value would be assigned an ordinal value of 5.
For this reason, the coefficients for marsh and saline wetland type vary between the two models.

All estimates of the factor loading (\ref{eqn:Z}) $\boldsymbol{\omega}$ are positive indicating that the linear relationship between latent wetland condition and each of the individual metrics is positive (Tables \ref{table:estimatesD} and \ref{table:estimatesDC}).
Based on the $95\%$ credible intervals these estimates are all significantly different from 0. 

The estimates of effective range of spatial correlation for the two models are comparable at 88 and 67 km.
The overall maximum distance between the 232 observed locations is $d_{max} = 470$ km whereas the maximum distance within the North Platte and Rio Grande River Basins is 93 km and 202 km, respectively.
The minimum distance between sampled locations from the two river basins is 240 km.
Not surprisingly, the estimate of the effective range indicates that the spatial correlation of wetland condition is only of interest within the river basins and not between them.

\begin{table}
\caption{Posterior estimates and 95\% credible interval for discrete-only model parameters.}
\begin{center}
\begin{tabular}{llrr}
\hline                     
Parameter& & Estimate &  ~~~~~~~~~~95 \% CI \\  
\hline                    
$\beta_1$ & Elevation   & 0.54 &(0.22, 0.89)\\
$\beta_2$ & Closed tree canopy   & 0.40&(0.20, 0.62)\\
$\beta_3$ & Saline   & 0.62 &(0.16, 1.10)\\
$\beta_4$ & Marsh   & 0.60 &(0.26, 0.97)\\
$\beta_5$ & Wet meadow  & -0.03 &(-0.28, 0.21)\\
$\beta_6$ & Fen & 1.00 &(0.55, 1.53)\\   \hline  
$3/\phi_2$&Effective Range & 0.88 &(0.45, 1.84)\\ \hline 
$\omega_1$& Native plant cover & 1.00&\\
$\omega_2$ & Noxious weed cover & 1.37 &(1.00, 1.90) \\
$\omega_3$ & Aggressive native cover & 2.54 &(0.89, 6.01) \\
$\omega_4$ & Structural diversity & 0.21 &(0.11, 0.33) \\
$\omega_5$ & Floristic quality & 1.59 &(1.33, 1.91)  \\ \hline
$\sigma^2_1$ &Native plant cover & 1.00&\\
$\sigma^2_2$ & Noxious weed cover & 1.34 &(0.86, 2.16)\\
$\sigma^2_3$ & Aggressive native cover & 20.46 &(8.00, 67.64)\\
$\sigma^2_4$ & Structural diversity & 0.89 &(0.67, 1.18)\\
$\sigma^2_5$ & Floristic quality & 0.36 &(0.22, 0.57)\\ 
\hline     
\end{tabular} 
\end{center}
\label{table:estimatesD}
\end{table}

\begin{table}
\caption{Posterior estimates and 95\% credible interval for mixed response model parameters.}
\begin{center}
\begin{tabular}{llrr}
\hline                     
Parameter && Estimate & ~~~~~~~~~~95 \% CI \\  
\hline                    
$\beta_1$ & Elevation   & 0.39 &(0.23, 0.57)\\
$\beta_2$  & Closed tree canopy  & 0.17 &(0.07, 0.28)\\
$\beta_3$ &Saline   & -0.21 &(-0.51, 0.07)\\
$\beta_4$  &Marsh  & -0.22 &(-0.43, -0.03)\\
$\beta_5$   &Wet meadow & -0.26 &(-0.42, -0.12)\\
$\beta_6$  & Fen  & 0.22 &(0.05, 0.42)\\   \hline  
$3/\phi_2$&Effective Range & 0.67 &(0.31, 3.08)\\ \hline 
$\omega_1$ & Native plant cover &1.00&\\
$\omega_2$ & Noxious weed cover &1.21 &(0.86, 1.69) \\
$\omega_3$ & Aggressive native cover &5.37 &(2.28, 10.57) \\
$\omega_4$ & Structural diversity & 0.38 &(0.25, 0.54) \\
$\omega_5$ &Floristic quality & 1.52 &(1.28, 1.83)  \\ \hline
$\sigma^2_1$ &Native plant cover &1.00&\\
$\sigma^2_2$ & Noxious weed cover &1.36 &(0.89, 2.18)\\
$\sigma^2_3$ & Aggressive native cover &11.83 &(4.41, 33.67)\\
$\sigma^2_4$ & Structural diversity & 0.61 &(0.46, 0.81)\\
$\sigma^2_5$ &Floristic quality &  0.18 &(0.13, 0.24)\\ 
\hline     
\end{tabular} 
\end{center}
\label{table:estimatesDC}
\end{table}

To compare the performance of the discrete-only model to the mixed response model, we compute the median squared error loss using the latent response $\mathbf{Z}$. 
For the ordinal metrics, we estimate the ``true" value of $\mathbf{Z}$ using (\ref{eqn:fitZ}). 
The squared error loss for each metric is similar between the discrete-only model and the mixed response model (See Table \ref{table:loss} in Appendix \ref{sec:Aloss}).

The remaining results presented here are for the discrete-only model because it of interest to the ecologists.
The multiple correlation statistics (\ref{eqn:R}) suggest that metric 5, floristic quality assessment, is most closely correlated with wetland condition (Table \ref{table:rhoD}) and should be ranked most important in evaluating wetland condition.
The assessments of native plant cover, noxious weed cover, and aggressive native cover are moderately related to wetland condition.
The structural diversity measurement (metric 4) appears to be the least correlated with wetland condition of the five measurements and therefore is ranked last.
Estimates of percent contribution are also given in Table \ref{table:rhoD} where the values are calculated based on the estimate of $R_{Z_j|H}$ divided by the sum of all estimates of $R_{Z_j|H}$ for $j = 1, \dots, 5$.
The percent contribution estimates can be used as weights for each of the metrics in estimating the underlying wetland condition.
The last column in Table \ref{table:rhoD} reports the current index weights that were selected by a group of wetland experts \citep{Lemly2012}.
The scientists believe floristic quality assessment to be the most important. 
The weight ``$0$ or $20\%"$ assigns $20\%$ weight to the lower of the noxious weed cover and aggressive native cover metrics.
Our estimates improve on the current weighting scheme by being statistically derived weights for each of the metrics with confidence limits.

\begin{table}
\hspace{-.2cm}
\footnotesize{
\caption{Discrete-only model: For each metric, estimates and 95\% credible intervals for the multiple correlation value, estimates of the percent contribution, and rank in evaluating wetland condition.}
\begin{center}
\begin{tabular}{l|c|cc|cc|c|r}
\hline
Metric & Parameter & ~Est.~~~~~ & 95 \% CI & \% Contrib. & 95 \% CI  &Rank & Index\\
\hline
Native plant cover &$R_{Z_1|H}$ &0.80& (0.68,  0.88)&0.23&(0.20,  0.26)&3 &20\%\\
Noxious weed cover &$R_{Z_2|H}$&0.84& (0.70, 0.92)&0.24&(0.21,  0.28)&2&0 or 20\%\\
Aggressive native cover~&$R_{Z_3|H}$&0.58& (0.23,  0.83)&0.17&(0.08,  0.22)&4&0 or 20\%\\
Structural diversity&$R_{Z_4|H}$&0.28& (0.09,  0.49)&0.08&(0.03, 0.13) &5&20\%\\ 
Floristic quality&$R_{Z_5|H}$&0.96& (0.92,  0.98)&0.28&(0.25,  0.32) &1&40\%\\ 
\hline
\end{tabular}
\end{center}
\label{table:rhoD}}

\end{table}

\begin{figure}
\begin{center}
\includegraphics[height=1.7in]{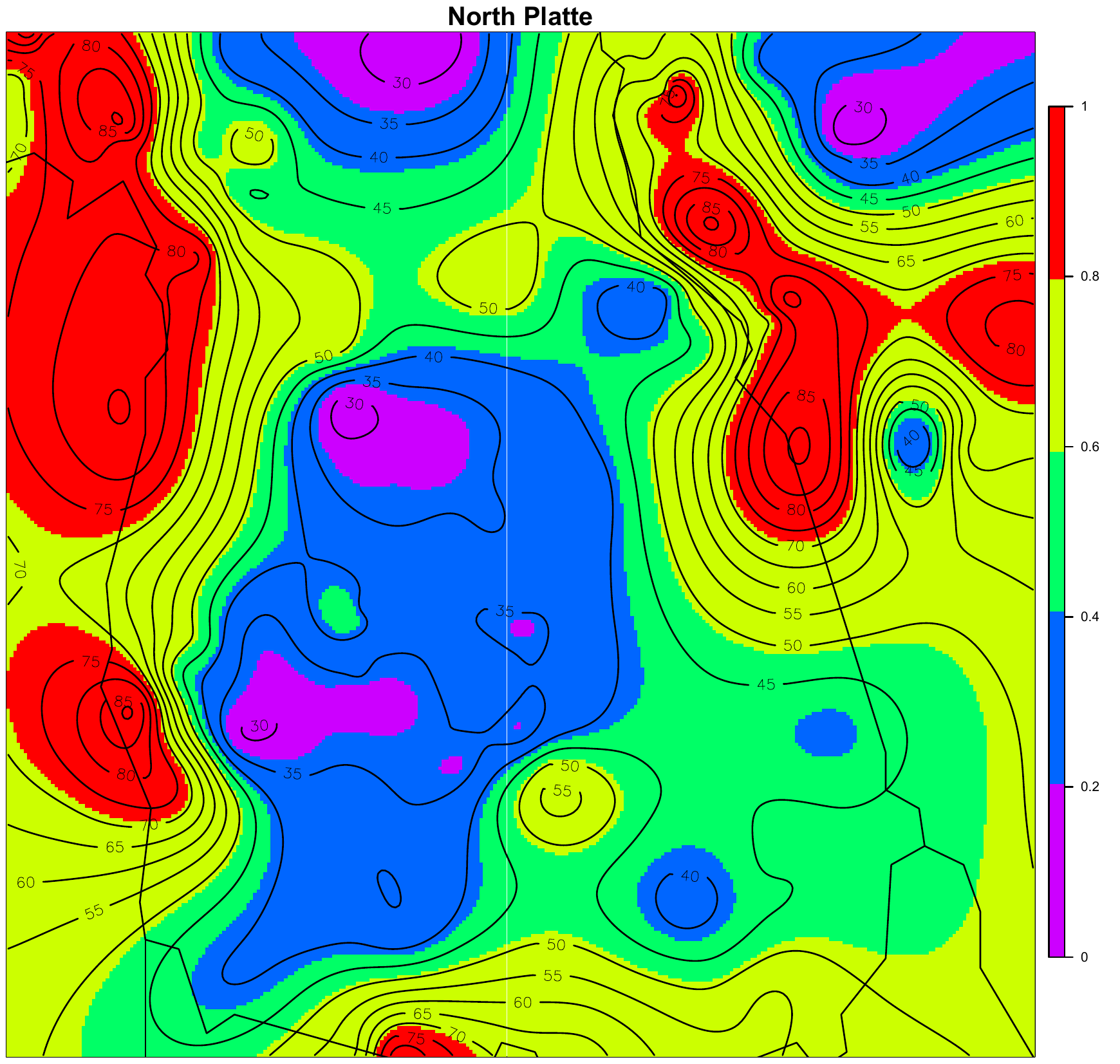}
\hspace{.1in}
\includegraphics[height=1.7in]{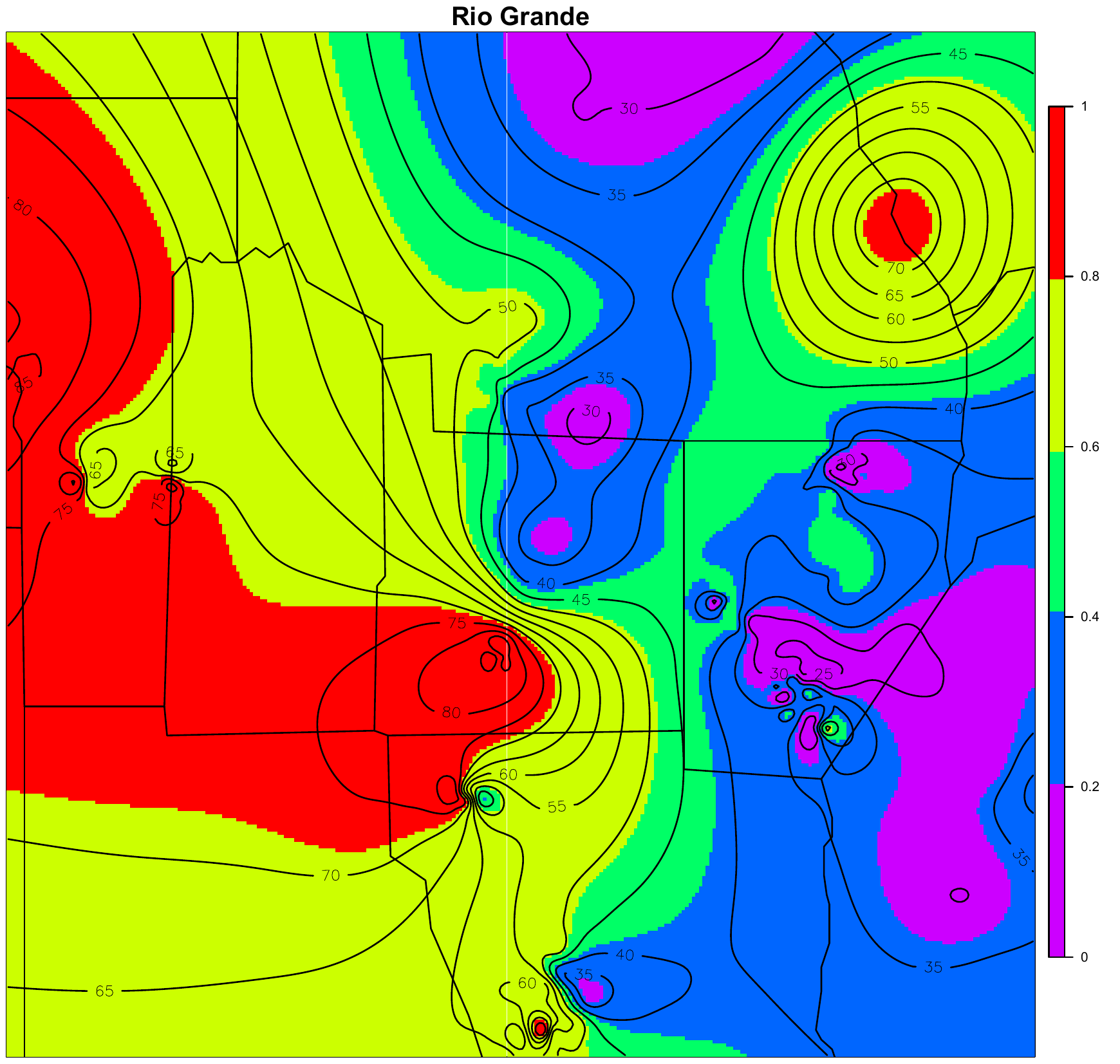}
\caption{Median of the posterior ranks of the latent spatial process encompassing wetland condition ($\mathbf{H}$) across space from the discrete-only response model.}
\label{Hsurf}
\end{center}
\end{figure}

We estimate the latent spatial process $\mathbf{H}$ of wetland condition within the North Platte and Rio Grande River Basins by drawing from the posterior distribution $p({H}_i|\mathbf{y})$  for $i = 1, \dots, n$.  
Since the values of ${\mathbf{H}}$ hold no intrinsic value, we rank the locations from draws of the posterior distribution. 
For each draw $t$ in $1, \dots, T$, the posterior value of $H_i$ is ranked across all $i = 1, \dots, n$ assigning a posterior rank to each location for each draw.
We estimate the latent spatial process of wetland condition by computing the median of the posterior ranks at each location.  
A location with a median posterior rank falling in the top 20\% of ranks indicates that the wetland at this particular location is in the top 20\% of all wetlands in the region in terms of biotic condition.
Figure \ref{Hsurf} shows the median of the posterior ranks across all locations within the North Platte and Rio Grande River Basins. 
Linear interpolation is used to provide a relatively smooth surface over the two river basins. 
Note, however, that wetlands are not found continuously over the regions. 
The color scale and contours of the surface are based on the percentile of the median of the posterior ranks over all locations. 
Wetland management efforts should be directed towards areas within the river basins with low posterior ranks. 
For example, the wetlands in the eastern region of the Rio Grande River Basin may be of concern. Conversely, land managers may wish to preserve wetlands in good condition such as those shown in red in Figure \ref{Hsurf}.
Similar plots can be made for the estimates of uncertainty. 
We performed a simulation study to evaluate the model and out-of-sample predictive performance (Appendix \ref{sec:sim}). 
The results indicate that our method provides accurate parameter estimates, predictions, and predictive coverage for the simulation scenarios that we considered.

\section{Discussion}
\label{sec:Disc}
The multilevel multivariate latent Gaussian process model presented in this paper provides a method for evaluating a continuous latent Gaussian process using mixed ordinal and continuous multivariate response data.
A multivariate latent variable is used as the continuous representation of the multivariate mixed response.
A second latent variable depending on site-specific covariates models the continuous random field that is assumed to be driving the multivariate response.
Whereas the continuous latent random field was modeled in this work using a Gaussian process, \cite{LindgrenRue11} present an approximation to the Gaussian field using a Gaussian Markov random field.
This approach could accelerate estimation of the parameters of the spatial covariance function.

Our multilevel multivariate latent variable model is used to evaluate the ecological condition of wetlands or other natural resources.
Whereas \cite{Liu2005} gave a general framework for spatial structural equation modeling, the model presented here for multivariate response data could be easily replicated or modified for other applications. 
The model is advantageous as it allowed for comparisons of the condition of wetlands in two river basins in Colorado across space.
Further, in-field measurements, or metrics, were ranked when evaluating the wetland condition score at each particular location.
These rankings allow assignment of statistically valid weights to the five measurements or metrics.
These results will lead to a decrease in the time and effort needed for future wetland evaluation.
They will also help land managers to design and implement effective protocols for maintaining and restoring wetland habitats.

While we have described and applied the model to a problem related to wetland condition, the model holds in much larger context. 
For example, in human health, doctors apply a panel of tests to a subject to evaluate health. 
In this case, the multivariate response would be the outcomes of the tests and the covariates would be individual information such as gender and body mass index (BMI). 

\begin{acknowledgement}
This work was supported by U.S. Environmental Protection Agency (EPA-1605-09) awarded to the Colorado Natural Heritage program at Colorado State University.
Hoeting was also supported by the National Science Foundation (EF-0914489).
We send our special thanks to Joanna Lemly and Laurie Gilligan of the Colorado Natural Heritage Program without whom this work would not be possible.
We would like to thank the anonymous referees whose detailed comments improved this paper.
\end{acknowledgement}


\bibliographystyle{apalike}
\bibliography{biblioPhD}

\begin{thebibliography}{}

\bibitem[Albert and Chib, 1993]{AlbertChib93}
Albert, J. and Chib, S. (1993).
\newblock {B}ayesian analysis of binary and polychotomous response data.
\newblock {\em Journal of the American Statistical Association},
  88(422):669--679.

\bibitem[Albert and Chib, 1997]{AlbertChib97}
Albert, J. and Chib, S. (1997).
\newblock {B}ayesian methods for cumulative, sequential and two-step ordinal
  data regression models.
\newblock Technical report, Citeseer.

\bibitem[Chakraborty et~al., 2010]{Chakraborty10}
Chakraborty, A., Gelfand, A., Wilson, A., Latimer, A., and Silander~Jr, J.
  (2010).
\newblock Modeling large scale species abundance with latent spatial processes.
\newblock {\em The Annals of Applied Statistics}, 4(3):1403--1429.

\bibitem[Chiu et~al., 2011]{Chiu10}
Chiu, G., Guttorp, P., Westveld, A., Khan, S., and Liang, J. (2011).
\newblock Latent health factor index: a statistical modeling approach for
  ecological health assessment.
\newblock {\em Environmetrics}, 22(3):243--255.

\bibitem[Christensen and Amemiya, 2002]{Christensen02}
Christensen, W. and Amemiya, Y. (2002).
\newblock Latent variable analysis of multivariate spatial data.
\newblock {\em Journal of the American Statistical Association},
  97(457):302--317.

\bibitem[Dahl, 2011]{Dahl2011}
Dahl, T. (2011).
\newblock {\em Status and trends of wetlands in the conterminous United States
  2004 to 2009}.
\newblock US Dept. of the Interior, US Fish and Wildlife Service.

\bibitem[Gelman et~al., 2004]{Gelman2004BDA}
Gelman, A., Carlin, J., Stern, H., and Rubin, D. (2004).
\newblock {\em {B}ayesian data analysis}.
\newblock CRC press.

\bibitem[Higgs and Hoeting, 2010]{Higgs10}
Higgs, M. and Hoeting, J. (2010).
\newblock A clipped latent variable model for spatially correlated ordered
  categorical data.
\newblock {\em Computational Statistics \& Data Analysis}, 54(8):1999--2011.

\bibitem[Karr, 1981]{Karr81}
Karr, J. (1981).
\newblock Assessment of biotic integrity using fish communities.
\newblock {\em Fisheries}, 6(6):21--27.

\bibitem[Lemly and Gillian, 2012]{Lemly2012}
Lemly, J. and Gillian, L. (2012).
\newblock {\em North {P}latte wetland profile and condition assessment}.
\newblock Colorado {N}atural {H}eritage {P}rogram, {C}olorado {S}tate
  {U}niversity, {F}ort {C}ollins, {C}olorado.

\bibitem[Lemly et~al., 2011]{Lemly2011}
Lemly, J., Gillian, L., and Fink, M. (2011).
\newblock {\em Statewide strategies to improve effectiveness in protecting and
  restoring {C}olorado's wetland resource}.
\newblock Colorado {N}atural {H}eritage {P}rogram, {C}olorado {S}tate
  {U}niversity, {F}ort {C}ollins, {C}olorado.

\bibitem[Lindgren et~al., 2011]{LindgrenRue11}
Lindgren, F., Rue, H., and Lindstr{\"o}m, J. (2011).
\newblock An explicit link between {G}aussian fields and {G}aussian {M}arkov
  random fields: the stochastic partial differential equation approach.
\newblock {\em Journal of the Royal Statistical Society: Series B (Statistical
  Methodology)}, 73(4):423--498.

\bibitem[Liu et~al., 2005]{Liu2005}
Liu, X., Wall, M., and Hodges, J. (2005).
\newblock Generalized spatial structural equation models.
\newblock {\em Biostatistics}, 6(4):539--557.

\bibitem[Muthen, 1984]{Muthen1984}
Muthen, B. (1984).
\newblock A general structural equation model with dichotomous, ordered
  categorical, and continuous latent variable indicators.
\newblock {\em Psychometrika}, 49(1):115--132.

\bibitem[{R Development Core Team}, 2007]{R}
{R Development Core Team} (2007).
\newblock {\em R: A Language and Environment for Statistical Computing}.
\newblock R Foundation for Statistical Computing, Vienna, Austria.
\newblock {ISBN} 3-900051-07-0, {URL} http://www.R-project.org.

\bibitem[Rencher, 2002]{Rencher2002}
Rencher, A. (2002).
\newblock {\em Methods of Multivariate Analysis, Second Edition}.
\newblock John Wiley \& Sons, Inc.

\bibitem[Rocchio, 2007]{Rocchio2007}
Rocchio, J. (2007).
\newblock {\em Floristic quality assessment indices for {C}olorado plant
  communities}.
\newblock Colorado {N}atural {H}eritage {P}rogram, {C}olorado {S}tate
  {U}niversity, {F}ort {C}ollins, {C}olorado.

\bibitem[Schmidt et~al., 2008]{Schmidt2008}
Schmidt, A., Concei{\c{c}}{\~a}o, M., and Moreira, G. (2008).
\newblock Investigating the sensitivity of {G}aussian processes to the choice
  of their correlation function and prior specifications.
\newblock {\em Journal of Statistical Computation and Simulation},
  78(8):681--699.

\bibitem[Skrondal and Rabe-Hesketh, 2004]{SkrondalRabe04}
Skrondal, A. and Rabe-Hesketh, S. (2004).
\newblock {\em Generalized latent variable modeling: multilevel, longitudinal,
  and structural equation models}.
\newblock CRC Press.

\bibitem[Stevens and Olsen, 2004]{Stevens04}
Stevens, D. and Olsen, A. (2004).
\newblock Spatially balanced sampling of natural resources.
\newblock {\em Journal of the American Statistical Association}, 99(465).

\bibitem[Swink and Wilhem, 1994]{Swink1994}
Swink, F. and Wilhem, G. (1994).
\newblock Plants of the {C}hicago {R}egion. 4th edition.
\newblock {\em Indiana Academy of Science, Indianapolis, IN}.

\bibitem[Taft et~al., 1997]{Taft1997}
Taft, J., Wilhelm, G., Ladd, D., and Masters, L. (1997).
\newblock {\em Floristic quality assessment for vegetation in {I}llinois, a
  method for assessing vegetation integrity}.
\newblock Illinois Native Plant Society.

\end{thebibliography}

\newpage
\section{Appendix}
\label{sec:Appendix}
\addcontentsline{toc}{section}{Appendix}

\subsection{Observed Data}
\label{data}
\begin{figure}[h!]
\begin{center}
\includegraphics[height=3.5in]{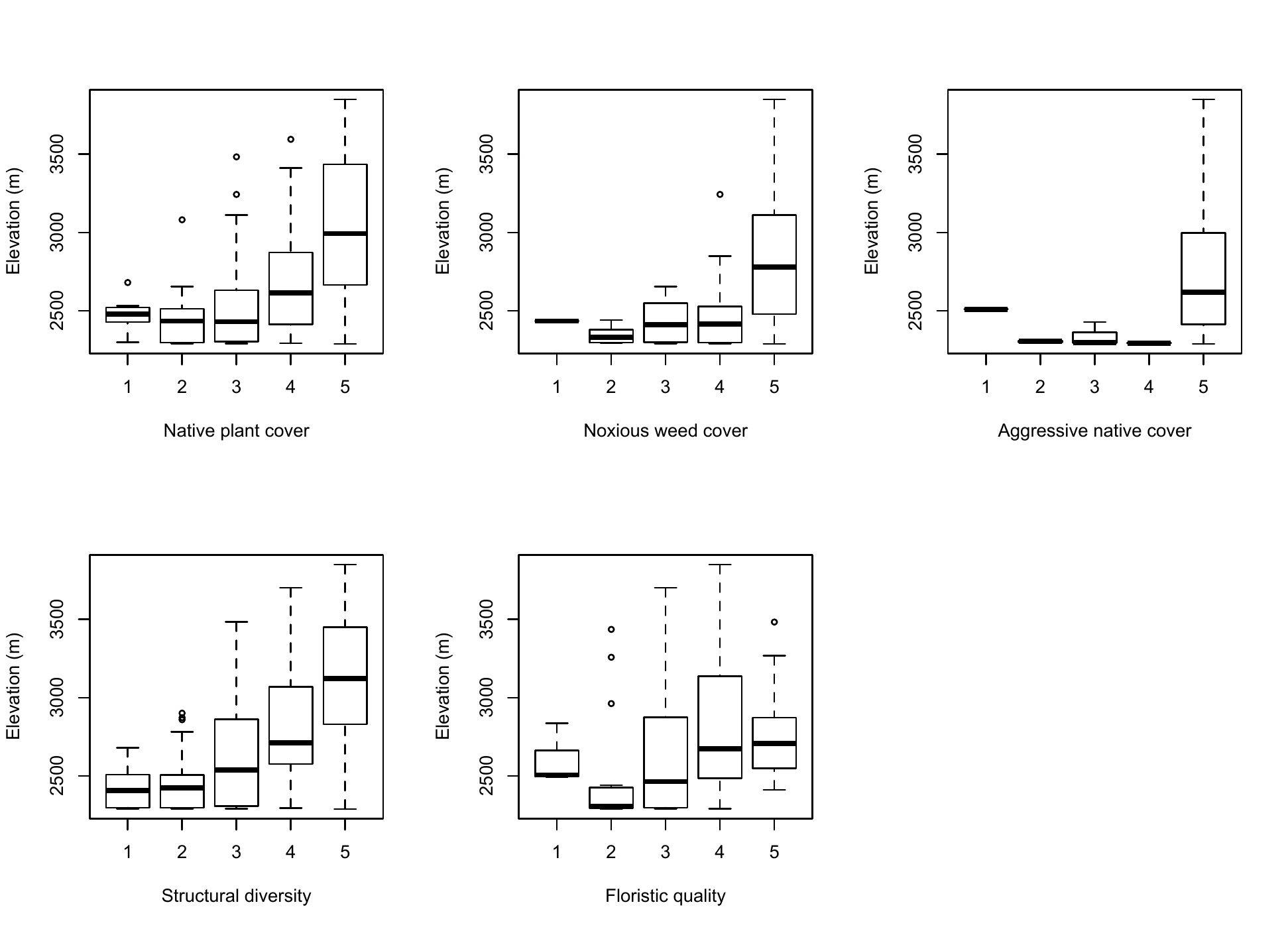}
\caption{Boxplots of elevation (y-axis) for each ordinal response (x-axis) for each metric.} 
\label{BoxElev}
\end{center}
\end{figure}

\begin{figure}[h!]
\begin{center}    
\includegraphics[height=3.5in]{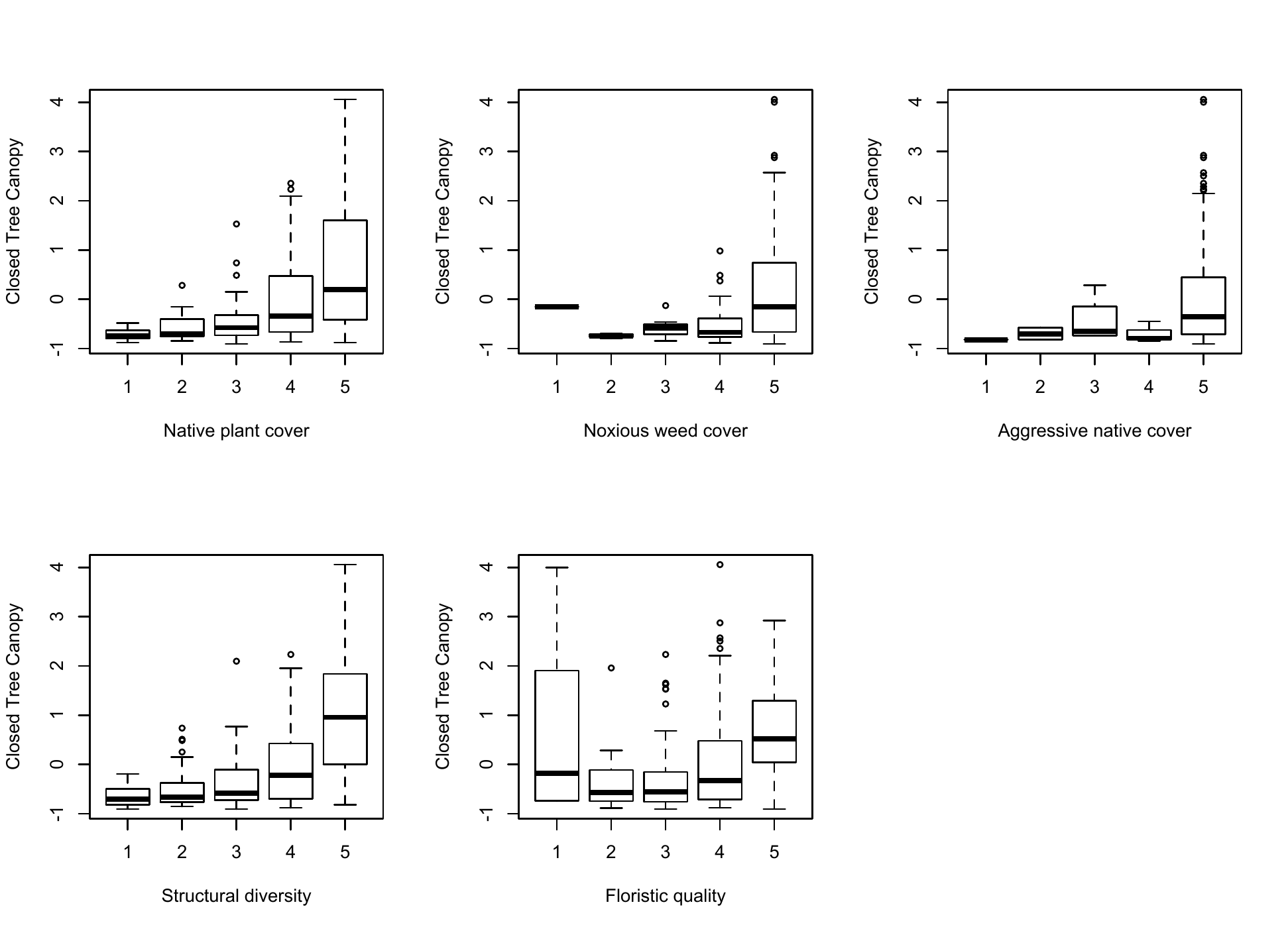}
\caption{Boxplots of closed tree canopy (y-axis) for each ordinal response (x-axis) for each metric.} 
\label{BoxCTC}
\end{center}
\end{figure}

The frequency of the observed ordinal response values for each metric over all $n=232$ locations are summarized in Table \ref{table:Raw}. Figures \ref{BoxElev} and \ref{BoxCTC} show univariate summaries between the each ordinal response and the covariates.

\begin{table}[h!]
\caption{Observed response data by metric}
\begin{center}
\begin{tabular}{l|ccccr}
\hline
& \multicolumn{5}{c}{Ordinal response}\\ \hline 
Metric & ~~~~1~~~~ & ~~~~2~~~~ & ~~~~3~~~~ & ~~~~4~~~~ & ~~~~5\\ \hline 
Native plant cover  & 9&16&60&69&75\\
Noxious weed cover & 1&6&10&50&165\\ 
Aggressive native cover~~ & 1&4&4&3&220\\ 
Structural diversity & 5&15&82&108&22\\ 
Floristic quality & 24&47&65&26&69\\ \hline
\end{tabular}
\end{center}
\label{table:Raw}
\end{table}

\subsection{Squared error loss}
\label{sec:Aloss}

The discrete-only model and the mixed response model are compared by computing the median squared error loss using the posterior predictions of the latent response $\mathbf{Z}$.
The ``true" value of $\mathbf{Z}$ for the ordinal metrics is estimated using (\ref{eqn:fitZ}). 
To compare squared error loss across models and metrics, we scale each loss value by the variance of its ``true" value of $\mathbf{Z}$. 
The standardized loss for each location $i$ and metric $j$ is computed using the posterior draws as
\begin{equation}
\label{eq:LossStand}
\frac{(Z_{ij}^{(m)}-\hat{Z}_{ij})^2}{\hat{\sigma}^2_{\hat{\mathbf{Z}}_{j}}}
\end{equation}
where $Z_{ij}^{(m)}$ is the $m^{th}$ draw of $Z_{ij}$ and $\hat{Z}_{ij}$ is the true value of the continuous representation of the observed ordinal response, $Y_{ij}$. 
For a continuous response metric, $\hat{\sigma}^2_{\hat{\mathbf{Z}}_{j}}$ is the variance of the response vector $\boldsymbol{Y}_j$ since $\hat{\mathbf{Z}}_j$ is observed.
 For an ordinal response metric, $\hat{\sigma}^2_{\hat{\mathbf{Z}}_{j}}$ is the variance of $\hat{\mathbf{Z}}_{j}$, which is based on the MCMC draws. 
The resulting loss for each metric is similar between the discrete-only model and the mixed response model (Table \ref{table:loss}).

\begin{table}[h!]
\caption{Median squared error loss comparison between the two models.}
\begin{center}
\begin{tabular}{lcc}
\hline
& \multicolumn{1}{c}{~~Discrete-Only Model~~}& \multicolumn{1}{c}{~~Mixed Response Model~~}\\                    
Metric & Loss Estimate & Loss Estimate  \\  
\hline                    
Native plant cover  & 0.73 & 0.75\\
Noxious weed cover & 0.74 & 0.83 \\
Aggressive native cover & 1.70 & 1.04\\
Structural diversity & 0.96 & 0.93 \\ 
Floristic quality & 0.58 & 0.37 \\
\hline     
\end{tabular} 
\end{center}
\label{table:loss}
\end{table}

\subsection{Simulation Study}
\label{sec:sim}

\begin{figure}
\begin{center}
\includegraphics[height=4in]{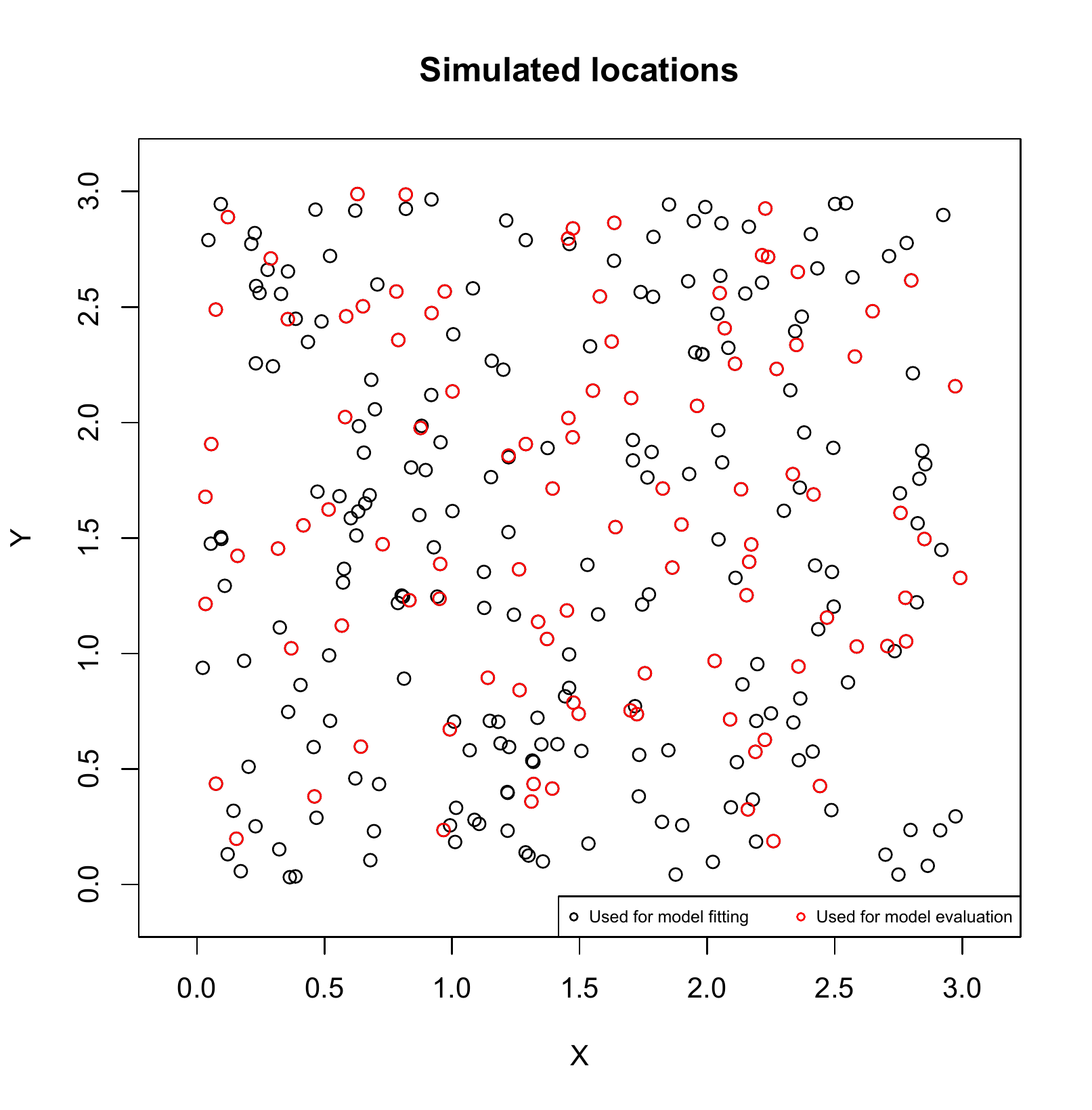}    
\caption{Three hundred simulated locations within 3 $\times$ 3 grid. Two hundred locations used to fit the model and the remaining one hundred were used for model evaluation.} 
\label{SimLocs}
\end{center}
\end{figure}

To evaluate the performance of our methods we simulated data based on the multivariate multilevel latent variable model with three discrete response metrics.
Three datasets were simulated as outlined below and the model was estimated for each dataset. 
The outcomes of the three simulation were similar and therefore we report the results of only one. 
We define our spatial domain of interest as a $3 \times 3$ spatial grid and simulated 300 locations uniformly over the region.
We use the first $n=200$ locations to fit the proposed model and the remaining $m=100$ locations for prediction (Figure \ref{SimLocs}). 

We randomly simulated two multivariate random variables over the spatial domain to use as covariates $\mathbf{X_1}$ and $\mathbf{X_2}$ in the mean of $\mathbf{H}$ where $\mathbf{H}$ is drawn according (\ref{Hdist}). 
Values for the coefficient vector $\boldsymbol{\beta}=(\beta_1, \beta_2)$ were fixed at 0.22 and 0.95, respectively.
The true parameter of the covariance of $\mathbf{H}$ was fixed at $\phi_1 = 1$ and $\phi_2 = 15.76$.
Note that $\phi_1$ is the sill parameter of the covariance and $\phi_2$ is the range parameter of the exponential correlation function. 
The effective range of the exponential correlation function is $3/ \phi_2 = 0.19$. 
In this simulation, the spatial correlation is only large at locations at close proximity. 
The true latent spatial random variable $\mathbf{H}$ was a random drawn from a multivariate normal distribution with mean $\mathbf{X}\boldsymbol{\beta}$ and covariance $\Sigma_H(\boldsymbol{\phi}) = \phi_1 \exp^{-\mathbf{d}\phi_2}$ where $\mathbf{d}$ is the $n \times n$ matrix of distances between sample locations.
The true fixed effects $\boldsymbol{\theta}$ and $\boldsymbol{\omega}$ were randomly drawn from independent uniform and normal distributions, respectively where $\boldsymbol{\theta}=\{1.73, 3.80, 3.62\}$ and $\boldsymbol{\omega}=\{1.00, 1.37, -0.77\}$. 
The true values of $\theta_1$, $\theta_2$, and $\theta_3$ were all chosen to be positive to ensure that the observed ordinal response values spanned each of the $K = 5$ categories. 
For $j = 1, 2, 3$, the true continuous latent random vector $\mathbf{Z}_j$ was drawn from its multivariate normal distribution with mean $\theta_j \mathbf{1} + \omega_j \mathbf{H}$ and variance $\sigma_j^2 \mathbf{I}_n$. 
The length 6 vector of threshold values $\boldsymbol{\lambda}$ was fixed such that $\lambda_0=-\infty$, $\lambda_1 =0$, and $\lambda_5=\infty.$ 
The other thresholds were drawn from a multivariate normal distribution on the transformed scale and then were back transformed. 
This was done to preserve the constraint that $\lambda_k \le \lambda_{k+1}$. 
The resulting true threshold was set to \[\boldsymbol{\lambda}=\{-\infty, 0, 1.81, 3.26, 4.71, \infty\}.\] 
The observed ordinal response data $\mathbf{Y}_j$ for metrics $j = 1, \dots, 3$ are in the set $\{ 1, \dots, 5\}$ based on the values of $\mathbf{Z}_j$ and the true threshold vector $\boldsymbol{\lambda}$. 

We ran the MCMC algorithm for 100,000 iterations and discarded the first 10,000 as burn-in.
The true parameter values as well as the posterior median and 95\% credible intervals are given in Table \ref{table:SimulatedEstimates}.
The results show that all but two parameters, $\omega_2$ and $\sigma^2_2$ are captured their respective credible interval.


\begin{table}[!h]
\caption{Simulated parameter values and posterior median estimates and 95\% credible interval from model output.}
\begin{center}
\begin{tabular}{lrrr}
\hline                     
Parameter & ~~~~True Value &~~~Estimate &~~~~~~~~~~95 \% CI \\  
\hline                    
$\beta_1$    & 0.22 &0.21 &(0.04, 0.39)\\
$\beta_2$    & 0.95 &1.15 & (0.90, 1.41)\\   \hline  
$3/\phi_2$ & 0.19 &0.31 &(0.17, 0.66)\\ \hline 
$\omega_1$  &1.00& 1.00 &\\
$\omega_2$ & 1.37 &0.73&(0.47, 1.12) \\
$\omega_3$ &-0.77 &-0.80&(-1.01, -0.61)  \\ \hline
$\theta_1$ & 1.73&1.44& (0.73, 2.13)\\
$\theta_2$ & 3.80 &4.16&(3.30, 4.93) \\
$\theta_3$ &3.62 &4.01&(3.28, 4.83)  \\ \hline
$\sigma^2_1$ &1.00&1.00 &\\
$\sigma^2_2$ & 2.75 &0.96& (0.48, 1.97)\\
$\sigma^2_3$ &  1.41 &1.52& (1.03, 2.18)\\ 
\hline     
\end{tabular} 
\end{center}
\label{table:SimulatedEstimates}
\end{table}

Using the posterior draws of the model parameters, we make predictions using the Bayesian posterior prediction distributions $p(\widetilde{\mathbf{Y}}|\mathbf{y})$ and $p(\widetilde{\mathbf{H}}|\mathbf{y})$. 
We evaluated the predictive ability of the model by comparing the mode of the posterior prediction distribution to the true metric value at each site for each metric (Table \ref{table:simfits}). 
Of the 300 predicted metric scores, the truth was captured 57\% of the time, whereas the predicted metric value was within 1 of the truth 93\% of the time.

\begin{table}[h!]
\caption{The posterior modes and true discrete metric response values at the $m=100$ new locations for all metrics. In bold are the number of correct predictions of metric response values. }
\begin{center}
\begin{tabular}{c|ccccr}
\hline
&\multicolumn{5}{c}{True Value}\\ \hline
Posterior Median & ~~~~1~~~~ & ~~~~2~~~~ & ~~~~3~~~~ & ~~~~4~~~~ & ~~~~5\\ \hline
       1 &$\mathbf{1}$ &1&  0  &0 & 0\\ 
       2 &10 &$\mathbf{15}$  &8  &3  &0\\ 
       3 & 5  &29  &$\mathbf{36}$  &25  &3\\ 
       4  &0 &2  &4 &$\mathbf{15}$& 10\\ 
       5  &1 & 2  &5 &22 &$\mathbf{103}$\\ \hline
\end{tabular}
\end{center}
\label{table:simfits}
\end{table}

\begin{figure}[h!]
\begin{center}
\includegraphics[height=4in]{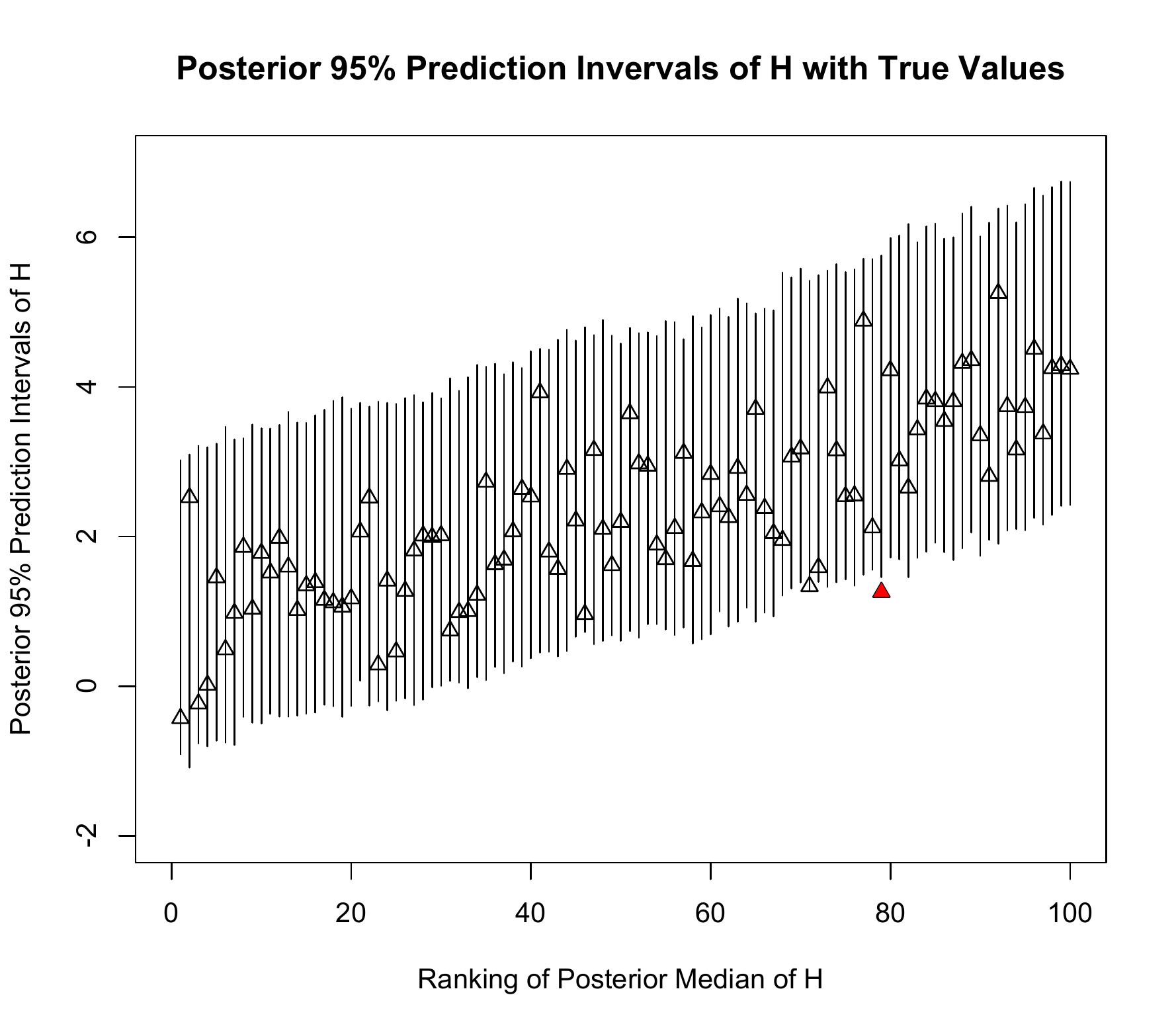}    
\caption{95\% posterior prediction intervals for latent spatial field $\mathbf{H}$ at $m = 100$ new locations. The true value is captured in 99 of the intervals.}
\label{PredictH}
\end{center}
\end{figure}

Capturing the latent random field $\mathbf{H}$ is of primary focus in this work. 
We make predictions of $\widetilde{\mathbf{H}}$ at $m=100$ new locations by taking draws from the Bayesian posterior prediction distribution $p(\widetilde{\mathbf{H}}|\mathbf{y})$.
95\% posterior prediction intervals of $\mathbf{\widetilde{H}}$ at $m=100$ new locations indicate that only one interval fail to capture the true value (Figure \ref{PredictH}).  
This indicates that our method achieves appropriate predictive coverage.

\end{document}